\documentstyle[preprint,aps,epsf,eqsecnum,tighten]{revtex} 
\preprint{\vbox{\baselineskip=12pt 
\rightline{CGPG-96/3-7} 
\rightline{gr-qc/9703037}}} 

\def\be{\nopagebreak[3]\begin{equation}} 
\def\ee{\end{equation}} 
\def\ba{\nopagebreak[3]\begin{eqnarray}} 
\def\ea{\end{eqnarray}} 
 
\def\ni{\noindent} 

\def\a{\alpha} 
\def\b{\beta} 
\def\c{\gamma} 
\def\d{\delta} 
\def\e{\eta} 
 
\def\f{\phi}

\def\l{\lambda} 
\def\m{\mu} 
\def\n{\nu} 
\def\o{\omega} 
\def\p{\pi}

\def\x{\xi}

\def\C{\Gamma}
\def\D{\Delta}

\def\L{\Lambda}

\def\S{\Sigma}

\newcommand{\teta}{\rlap{\lower2ex\hbox{$\,\tilde{}$}}\eta{}}
\newcommand{\tiN}{\rlap{\lower2ex\hbox{$\,\tilde{}$}}N{}}

\newcommand{\amodg}{\overline{\cal A/G}}
\newcommand{\AG}{\overline{\cal A/G}}

\newtheorem{theorem}{Theorem} 
\newtheorem{lemma}{Lemma} 

\begin{document}
\draft
\title{
A Combinatorial Approach to Diffeomorphism\\
Invariant Quantum Gauge Theories
}
\author {
Jos\'e A. Zapata\thanks{zapata@phys.psu.edu}
}
\address{
Center for Gravitational Physics and Geometry \\
Department of Physics, 
The Pennsylvania State University \\
104 Davey Laboratory, University Park, PA 16802
}
\maketitle
\begin{abstract} 
Quantum gauge theory in the connection representation 
uses functions of holonomies as configuration observables. 
Physical observables (gauge and diffeomorphism invariant) 
are represented in the Hilbert space of physical states; physical 
states are gauge and diffeomorphism invariant distributions on the space 
of functions of the holonomies of the edges of a certain family of graphs. 
Then a family of graphs embedded in the space manifold 
(satisfying certain properties) induces a 
representation of the algebra of physical observables. 
We construct a quantum model from the set of piecewise linear graphs 
on a piecewise linear manifold, and another manifestly combinatorial 
model from graphs defined on a sequence of increasingly refined 
simplicial complexes. Even though the two models are different at 
the kinematical level, they provide unitarily equivalent representations 
of the algebra of physical observables in 
{\em separable} Hilbert spaces of physical states 
(their s-knot basis is countable). Hence, the combinatorial framework 
is compatible with the usual interpretation of quantum field theory. 
\end{abstract}
\clearpage 

\section{Introduction} 

Quantum gauge theories can be described using the holonomies along the 
edges of a regular lattice as basic configuration observables. 
This idea was introduced by Wilson \cite{wilson} in the '70s and is now 
the basis of the modern lattice gauge theory. 
In diffeomorphism invariant gauge theories 
(like gravity using Ashtekar variables \cite{l,ashtekar} 
or Yang-Mills coupled to gravity) , the use of 
Wilson loops as primary observables of the theory led to the 
discovery of an interesting relation between quantum gauge theories 
and knot theory \cite{lee-carloLOOP}. 

Twenty years after the early works, the notion of Wilson loops was 
extended and serves as a rigorous foundation of quantum gauge field 
theory \cite{2dYMSU2}. The modern approach rests on the following idea: 
Begin by considering ``the family of all the possible lattice gauge 
theories'' defined on graphs whose edges are embedded in the base space. 
Then use a projective structure to organize the repeated information 
from graphs that share edges. 
For a manageable theory, the precise definition 
of ``the family of all the possible lattice gauge theories'' had to 
avoid situations where two different edges intersect each other an infinite 
number of times. The first solution to this problem \cite{analytic} 
led to the framework referred in this article as the 
analytic category; by restricting the set of 
allowed graphs $\C_\o $ to contain only graphs with 
piecewise analytic edges, 
one acquires a controllable theory. In the analytic 
category the diffeomorphisms are restricted to be analytic accordingly. 
After a subtle analysis, 
it was possible to sacrifice part of the simplicity of the results of the 
analytic case and extend the theory to the smooth category \cite{smooth}. 

While the foundations were solidifying, the theory also produced its first 
kinematical results for quantum gravity (the canonical 
quantization of gravity expressed in terms of Ashtekar variables). 
Regularized expressions for operators measuring 
the area of surfaces and volume of regions were developed \cite{geoperators}. 
These operators 
were also diagonalized and its eigenvectors were found to be labeled by 
spin networks (one-dimensional objects). In other words, a picture of 
polymer-like geometry arises from quantum gravity \cite{polymer}. 
A polymer-like geometry is predicted from a theory whose foundations 
require space to be an analytic manifold. This peculiar 
situation 
was the main motivation for the work presented in this article. 

In this article we present two quantum models: 
the combinatorial and the piecewise linear (PL) categories. 
The intention is to keep a simple framework that minimizes background 
structure and is suited to a polymer-like geometry, but 
that can still recover the classical macroscopic theory. 
Both models are based on the projective techniques used 
for the analytic and smooth categories; again, the difference relies on the 
family of graphs $\C$ considered and the corresponding ``diffeomorphisms.'' 

In the the piecewise linear category we fix a piecewise linear 
structure in the space manifold to specify the elements of the family 
of graphs $\C_{\rm PL}$ that define the Hilbert space. A piecewise 
linear structure on a manifold $\S$ can be specified by a division of the 
manifold into cells with a fixed affine structure (flat connection). Also 
it can be specified by a triangulation, that is, a fixed 
homeomorphism $\varphi : \S \to \S_0$ where $\S_0 \subset R^{2n+1}$ 
is a n-dimensional polyhedron with a fixed decomposition into simplices. 
An element of $\C_{\rm PL}$ is a graph 
whose edges are piecewise linear according to the fixed PL structure. 
This seems to be far from a background-free situation, 
but a PL structure is much weaker than an analytic structure; 
the same PL structure can be specified by any refinement of the 
original triangulation. 
Furthermore, we will prove that 
in three (or less) dimensions different choices of PL 
structures yield unitarily equivalent representations of the algebra 
of physical observables. 
This result is of particular interest for $3+1$ ($2+1$ or $1+1$) 
quantum models of pure gravity or of gravity coupled to Yang-Mills fields. 
To avoid confusion, we stress that the piecewise linear 
spaces used in this approach are not directly related to the ones used in 
Regge calculus. In simplified theories of gravity, like $2+1$ gravity and 
$BF$ theory, the lattice dual to the one induced by one of our piecewise 
linear spaces can be successfully related to a Regge lattice \cite{2+1-BF}. 
On the other hand, our approach contains a treatment based in cubic lattices 
as a particular case; the difference with the usual lattice gauge theory is 
that the continuum limit is taken by considering every lattice instead of 
just one. 

The manifestly combinatorial model has two main ingredients: 
simplicial complexes that describe geometry in combinatorial fashion, 
and a refinement mechanism that makes it capable 
to describe field theories. 
If we use a simplicial complex as the starting point of our combinatorial 
approach, the resulting model would 
be appropriate to describe topological field theories, but we want to generate 
a model for gauge theories with local degrees of freedom. A way to achieve 
this goal is to replace physical space (the base space) with a sequence of 
simplicial complexes $K_0, K_1, \ldots $ that are finer and finer. Our 
combinatorial model for quantum gauge theory is based in the family of 
graphs defined using our combinatorial representation of space. 

Even though the PL and the combinatorial categories 
are closely related, the resulting kinematical Hilbert spaces 
${\cal H}_{\rm kin_{PL}}$ and ${\cal H}_{\rm kin_C}$ are dramatically 
different. While the combinatorial Hilbert space ${\cal H}_{\rm kin_C}$ 
is separable 
(admits a countable basis), ${\cal H}_{\rm kin_{PL}}$ (like the Hilbert 
space constructed from the analytic category) is much bigger. 

Physically, what we need is a Hilbert space to represent 
physical (gauge and ``diffeomorphism'' invariant) observables; such 
Hilbert space can be constructed by ``averaging'' the states of the 
kinematic Hilbert space to produce physical states. 
An encouraging result is that 
the two models produce unitarily equivalent representations of the algebra 
of physical observables in the 
naturally isomorphic separable Hilbert spaces 
${\cal H}_{\rm diff_{PL}} , {\cal H}_{\rm diff_C}$. 
Separability in the combinatorial case is no 
surprise, and that both spaces of physical states (PL and combinatorial) 
are isomorphic follows from the fact that 
every knot-class of piecewise linear graphs has a representative that 
fits in our combinatorial representation of space. 

Two aspects of the loop approach to gauge theory are enhanced in its 
combinatorial version. 
On the mathematical-physics side, other approaches to quantum gravity 
coming from topological quantum field theory \cite{crane-lee} 
are much closer to the combinatorial category than they are to 
the analytic or smooth categories. 
On the practical side, the loop approach to quantum gauge theory is at least 
as attractive; a powerful computational technique comes built into 
this approach. Given any state in the Hilbert space of the continuum we can 
express it, to any desired accuracy, as a finite linear combination of 
states that come from the Hilbert space of a lattice gauge theory. 
Therefore, the matrix elements of every bounded operator can be computed, 
to any desired accuracy, in the Hilbert space of a lattice gauge theory. 
In this respect, the combinatorial picture presented in this article is 
favored because it is best suited for a computer implementation. 

We organize this article as follows. 
Section~\ref{review} reviews the general procedure to construct the
kinematical Hilbert space in the continuum starting 
from a family of lattice gauge 
theories. Then, in section~\ref{p-lspaces}, 
we carry out the procedure in the combinatorial and PL frameworks.  
In section~\ref{homeos}, we construct the physical Hilbert space. 
We treat separately the PL and combinatorial categories. Then we prove 
that the combinatorial and PL frameworks provide unitarily equivalent 
representations of the algebra of physical observables are unitarily 
equivalent. We also prove that the mentioned algebra of physical observables 
is independent of the background PL structure when the dimension of the space 
manifold is three or less. A summary, an analysis of some 
problems from the combinatorial perspective and a comparison with the analytic 
category are the subjects of the concluding section. 

\section{From Quantum Gauge Theory in the Lattice to the Continuum 
Via the Projective Limit: A Review} \label{review} 

A connection on a principal bundle is
characterized by the group element that it assigns to  
every possible path in the base space. 
Historically, this simple observation 
led to treat the set of holonomies for all 
the loops of the base space as the basic configuration 
observables to be promoted to operators. 

Now we start the 
construction of a kinematical Hilbert space for quantum gauge theories. 
To avoid extra complications, 
we only treat cases with a compact base space $\S$ and 
we restrict our atention to trivializable 
bundles over $\S$.  For convenience, we start with a fixed trivialization. 
In the modern approach (Baez, Ashtekar et al \cite{analytic}) 
the concept of paths or loops has been extended to that of 
graphs $\c \subset \S$ whose edges, in contrast with their 
predecessors, are allowed to intersect. 

A {\em graph} 
$\c$ is, by definition, a {\em finite} set $E_\c$ of oriented edges 
and a set $V_\c$ of vertices satisfying the following conditions: \\
\begin{itemize} 
\item $e \in E_\c$ implies $e^{-1} \in E_\c$. 
\item The vertex set is the set of boundary points of the edges.  
\item The intersection set of two different edges 
$e_1, e_2 \in E_\c$ ($e_1 \neq e_2 , e_1 \neq e_2^{-1}$) is a subset of the 
vertex set. 
\end{itemize} 

Generally an edge $e \in E_\c$, is considered to be an equivalence class of 
not self-intersecting curves, under orientation preserving 
reparametrizations. Formally, 
$e:=[ e^\prime (I) \subset \S ]$ such that $e^\prime (I )\approx I$, 
where we denoted the unit interval by $I=[0,1]$. 
Composition of edges $e, f$ is 
defined if they intersect only at the final point of the initial edge and the 
initial point of the final edge 
$e^\prime (I)\cap f^\prime (I) = e^\prime (1) = f^\prime (0)$. 
Then the composition is defined by 
$f\circ e:=[f^\prime \circ e^\prime (I)]$; and 
given an edge $e:=[ e^\prime ]$ the edge defined by paths with the opposite 
orientation is denoted by $e^{-1}:=[ e^{\prime -1} ]$. 

The idea of considering ``every possible path'' in the base space to 
construct the space of generalized connections has to be made precise. 
Different choices in the class of edges that form the 
family of graphs considered lead to the different 
categories --analytic, smooth, PL and combinatoric-- 
of this general approach to 
diffeomorphism invariant quantum gauge theories. We denote a generic family 
of graphs by $\C$, and the analytic, smooth and combinatoric families by 
$\C_\o ,\C_\infty$, $\C_{\rm PL}$ and $\C_C$. 

A connection on a graph 
assigns a group element to each of the $2N_1$ graph's edges. Therefore, 
we can identify the space of connections ${\cal A}_\c$ of graph $\c$ 
with $G^{N_1}$. An element $A \in {\cal A}_\c$ is represented by 
$(A(e_1),A(e_1^{-1})=A(e_1)^{-1},\ldots,A(e_{N_1}), 
A(e_{N_1}^{-1})=A(e_{N_1})^{-1})$, where $A(e_i) \in G$. 

The collection of the spaces ${\cal A}_\c$ for every graph 
$\c \in \C$ gives an {\em over-complete} description of 
the space of generalized connections in the category specified by $\C$. 
For example, $\C_\o$ determines the analytic category and 
$\C_C$ specifies the combinatoric category. 

It is possible to organize all the repeated information by means of a
projective structure. 
We say that graph $\c$ is a refinement of graph $\c^\prime$ 
($\c\geq \c^\prime$) if the edges of $\c^\prime$ are ``contained'' in 
edges of $\c$; more precisely, if $e\in \c^\prime$ then either 
$e=e_1$ or $e=e_1\circ \ldots \circ e_n$ for some 
$e_1, \ldots , e_n  \in \c$. 
Given any two graphs  
related by refinement $\c\geq \c^\prime$ there is a projection 
$p_{\c^\prime\,\c}:{\cal A}_{\c}\rightarrow{\cal A}_{\c^\prime}$ 
\be
(A(e_1),A(e_2),\ldots,A(e_{N_1}))
\stackrel{p_{\c^\prime \c}}{\longrightarrow}
 (A^\prime(e_1)=A(e_2)A(e_1),A^\prime(e_2),\ldots,A^\prime(e_{N_1}))
\ee
where $e=e_1\circ e_2$, $e\in \c^\prime$, $e_1,e_2 \in \c$. 

The projection map and the refinement relation have two properties 
that will allow us to define $\overline{\cal A}$ as ``the space of 
connections of the finest lattice.'' 
First, we can easily check that $p_{\c\,\c^\prime}\circ p_{\c^\prime\,
\c^{\prime\prime}}=p_{\c\,\c^{\prime\prime}}$. 
Second, equipped with the refinement relation ``$\geq$'', the set 
$\C$ is a partially ordered, directed set; i.e. for all $\c$, 
$\c^\prime$ and $\c^{\prime\prime}$ in $\C$ we have: 
\be
\c\geq \c\;\;\;;\;\;\;\;\c\geq \c^\prime 
\;\;\;{\rm and}\;\;\;\c^\prime\geq \c
\Rightarrow \c=\c^\prime\;;\;\;\;\;\c\geq 
\c^\prime\;\;\;{\rm and}\;\;\;
\c^\prime\geq \c^{\prime\prime}\Rightarrow \c\geq 
\c^{\prime\prime}\;;
\ee
and, given any $\c^\prime,\c^{\prime\prime}\in \C$, there exists
$\c\in \C$ such that
\be
\c\geq \c^\prime\;\;\;\;{\rm and}\;\;\;\;\c\geq \c^{\prime\prime}.
\ee
This last property, that $\C$ is directed, is the only non trivial 
property; it will be proved for the PL and the 
combinatoric categories in the next section. 
The {\em projective limit} of the 
spaces of connections of all graphs yields the space of 
{\em generalized connections} $\overline{\cal A}$ 
\be
\overline{\cal A}:=\left\{(A_\c)_{\c\in \C}\in 
\prod_{\c\in \C}{\cal A}_\c\;\;:\;\;\c^\prime\geq \c
\Rightarrow p_{\c\,\c^\prime}A_{\c^\prime}=A_{\c}\right\}.
\ee
That is, the projective limit is contained in the cartesian product of
the spaces of connections of all graphs in $\C$, subject to the
consistency conditions stated above. There is a canonical projection $p_\c$
from the space $\overline{\cal A}$ to the spaces ${\cal A}_\c$ given by,
\be
p_\c\;\;:\;\overline{\cal A} 
\rightarrow 
{\cal A}_\c,\;\;\;p_\c((A_{\c^\prime}) _{\c^\prime\in \C})
:=A_\c. 
\ee 
With this projection, functions $f_\c$
defined on the space ${\cal A}_\c$ can be pulled-back to 
${\rm Fun}(\overline{\cal A})$. 
Such functions are called {\em cylindrical functions}. The sup norm 
\be 
|| f || _\infty = \sup_{A\in {\cal A}_\c} | f(A) |
\ee 
can be used to complete the space of cylindrical functions. As  result we 
get the Abelian $C^*$ algebra usually denoted by 
${\rm Cyl}(\overline{\cal A})$; 
to simplify the notation, in the rest of the article we will denote this 
algebra by ${\rm Cyl}_\Box$, where 
$\Box =\o ,\infty ,{\rm PL}, C$ 
labels the family of graphs defining the space of cylindrical functions 
considered. 

The uniform generalized measure $\m_0: {\rm Cyl}_\Box \to C$, 
sometimes called the Ashtekar-Lewandowski measure, is 
induced in ${\cal A}$ by the uniform (Haar) measure on the spaces 
${\cal A}_\c= G^{N_1}$. Other gauge invariant measures are available; 
when they are diffeomorphism invariant they induce ``generalized 
knot invariants'' (see \cite{genknot}). 
Finally, we define the kinematical Hilbert space to 
be the completion of ${\rm Cyl}({\cal A})$ on the norm induced by 
the (strictly positive) generalized measure $\m_0$ 
\be
{\cal H}_{\rm kin}:=L^2(\overline{\cal A},d\m_0).
\ee

This construction yields a cyclic representation of the algebra of cylindrical 
functions, the so called {\em connection representation}. 
Given a function defined on a lattice $\c$, for example the 
trace of the holonomy $T_\a$ along a loop $\a$ contained in $\c$, 
the corresponding operator $\hat{T}_\a$
will act by multiplication on states $\Psi_\c \in {\cal H}_{\rm kin}$: 
\be
(\hat{T}_\a\cdot\Psi_\c)(\bar{A}):=T_\a(\bar{A})\Psi_\c(\bar{A}).
\ee 

A complete set of Hermitian momentum operators on the Hilbert space 
$L^2(G_e,d\m_{\rm Haar})$ of a graph with a single edge $e$ come from 
the left $L_e(f)$ and right invariant $R_e(f)$ vector fields on $G_e$ 
as labeled by $f \in {\rm Lie}(G_e)$. These momentum operators are 
compatible with the projective structure \cite{dgeoAL}; thus, the set of 
momentum operators 
\be 
X_{\a ,e}(f) = 
\left\{ \begin{array} 
{r@{\quad }l} 
L_e(f) & \hbox{if edge }e \hbox{ goes out of vertex }\a \\
-R_e(f) & \hbox{if edge }e \hbox{ comes into vertex }\a 
\end{array} \right. \label{x}
\ee 
is a complete set of Hermitian momentum operators on 
${\cal H}_{\rm kin}$ when we use the generalized measure $\m_0$. 
In regularized expressions of operators involving the triad, 
the place of the triad is taken by the vector fields $X$; therefore, the 
measure $\m_0$ incorporates the physical reality conditions. 


Our main goal is to construct a Hilbert space where we can represent the 
algebra of physical (gauge and diffeomorphism invariant) observables. 
Because it is custumary we will proceed in steps; in this section we 
deal with the issue of gauge invariance 
and in the next with that of diffeomorphism invariance.
If we had chosen to generate the space of states invariant under both 
symmetries simultaneously we would arrive at the same result. 

A finite gauge transformation takes the 
holonomy $A_{e_1}$ to $g(\a)A_{e_1}g(\b)^{-1}$ (where edge $e_1$ goes from 
vertex $\a$ to vertex $\b$). Then a quantum gauge transformation 
is given by the unitary transformation 
\be 
G(g) \Psi_\c(A_{e_1}, \ldots A_{e_n}) := 
\Psi_\c(g(\a)A_{e_1}g(\b)^{-1}, \ldots g(\m)A_{e_n}g(\n)^{-1})
\ee 
Gauge transformations are just generalizations of right and left translations 
in the group. This implies that they are generated by left and right invariant 
vector fields. Given a graph $\c$, $C_\a(f)$ generates 
gauge transformations at vertex $\a$. Therefore gauge invariance of 
$\Psi_\c=\Psi_\c(A_{e_1}, \ldots A_{e_n})$ at vertex $\a$ means 
that it lies in the kernel of the Gauss constraint 
\be
C_\a(f) \cdot \Psi_\c :=\sum_{e\to \a}X^I_{e}\cdot \Psi_\c=0 \quad ,
\ee 
where the sum is taken over all the edges $e$ that start at vertex $\a$. 
Because it is a real linear combination of the 
momentum operators (\ref{x}), the Gauss constraint is essentially 
self-adjoint on ${\cal H}_{\rm kin}$. 

We could construct the space of connections modulo gauge 
transformations of a graph ${\cal A}_\c / {\cal G}_\c$. 
Then, using the same projective machinery, we could construct the Hilbert 
space $L^2(\AG,d\n_0)$. It is easy to see that the space of 
gauge invariant functions of $L^2(\overline{\cal A},d\m_0)$ is naturally 
isomorphic to ${\cal H}^{\prime}_{\rm kin}=L^2(\AG,d\n_0)$ 
if the measure $\n_0$ is the one induced by $\m_0$. The space 
${\cal H}^{\prime}_{\rm kin}$ of gauge 
invariant functions is spanned by spin network states. Spin network 
states are cylindrical functions 
$S_{\vec{\c} , j(e), c(v)}(A)$ labeled by an {\em oriented graph} 
(a graph $\c$ plus a choice of either $e\in E_\c$ or 
$e^{-1} \in E_\c$, for every edge in $\c$, to belong to the 
oriented graph $\vec{\c}$) whose edges and vertices are colored. The 
``colors'' 
$j(e)$ on the edges $e\in E_\c$ assign a non trivial irreducible 
representation of the gauge group to the edges. 
And the ``colors'' $c(v)$ on the vertices 
$v\in V_\c$ assign to each vertex a 
gauge invariant contractor (intertwining operator) 
that has indices in the representations determined by the colored 
edges that meet at the vertex. The spin network states is defined by 
\be \label{S} 
S_{\vec{\c} , j(e), c(v)}(A) = 
\prod_{e\in E_{\vec{\c}}} \p _{j(e)}[A(e)] \cdot 
\prod_{ v\in V_{ \vec{\c} } } c(v) \quad , 
\ee 
where `$\cdot$' stands for contraction of all the indices of the 
matrices attached to the edges with the indices of the intertwiners 
attached to the vertices. In the inner product that the uniform measure 
$\m_0$ induces in ${\cal H}^{\prime}_{\rm kin}$ two spin network states are
orthogonal if they are not labeled by the same (unoriented) graph or 
if their edge's colors are diferent. 
For calculational purposes it is convenient to choose an orthonormal basis for 
${\cal H}^{\prime}_{\rm kin}$ by normalized spin network states with 
special labels of the intertwining operators assigned to the vertices; see 
\cite{rovelli-depietri}.

\section{PL and combinatoric categories}
\label{p-lspaces} 

In this section we construct two quantum models using the general framework 
outlined above. First the family of piecewise linear (PL) graphs is 
introduced. Then we prove that it is a partially ordered, directed set. 
As a result, the algebra of functions of the connection defined by the 
PL graphs has a cyclic representation in the Hilbert space 
${\cal H}_{\rm kin_{PL}}$. The second subsection briefly reviews some 
elements of combinatoric topology while constructing the 
family of combinatoric graphs. In this case, the resulting algebra of 
functions is represented in the separable Hilbert space 
${\cal H}_{\rm kin_C}$. While at this level the two quantum models 
yield completely different Hilbert spaces, in the section~\ref{homeos} 
we will 
prove that the corresponding spaces of ``diffeomorphism'' invariant states 
are naturally isomorphic. 

\subsection{The PL category} 
To specify the elements of the family 
of graphs $\C_{\rm PL}$ that define the Hilbert space of the PL 
category we need a fixed piecewise linear structure on space $\S$. 
A piecewise 
linear structure on a manifold $\S$ can be specified by a division of the 
manifold into cells with a fixed affine structure (flat connection). Also 
it can be specified by a triangulation, that is, a fixed 
homeomorphism $\varphi : \S \to \S_0$ where $\S_0$ is a n-dimensional 
polyhedron with a fixed decomposition into simplices. 
To be more explicit, we can use the fact that 
every n-dimensional polyhedron can be embedded in $R^{2n+1}$ and consider 
from the beginning $\S_0\subset R^{2n+1}$. 
Then $\S_0$ can be decomposed into a collection of convex cells 
(geometrical simplices). 
A {\em geometric simplex} in $R^{2n+1}$ is 
simply the convex region defined by its set of vertices 
$\{ {\bf s}_0, \ldots , {\bf s}_k \}$, ${\bf s}_i \in R^{2n+1}$ 
\be 
\D(\{ {\bf s}_0, \ldots , {\bf s}_k \} )= \{ {\bf s} = 
\sum_{i=0}^k t_i {\bf s}_i \} 
\ee 
\ni
where $t_i \in [ 0,1]$ and $\sum_{i=0}^k t_i =1$. 
The triangulation of $\S_0$ fixes an affine structure in its cells, 
namely, a PL structure. 
Using the local affine coordinate systems $t_i$, we can decide which curves 
are straight lines inside any cell. 
Then a piecewise linear curve in $\S_0$ is a curve that is straight inside 
every cell except for a {\em finite} set of points; in this set of points and 
in the points where it crosses the boundaries of the cells the curve bends, 
but is continuous. 

A piecewise linear graph $\c \in \C_{\rm PL}$ is a graph (according to  
the definition given in the previous section) such that every edge 
$e \in E_\c$ is piecewise linear. 

In the previous section we gave a natural partial 
order (``refinement relation'', $\geq$) for any family of graphs. 
Our task is now to prove that the partially ordered set $\C_{\rm PL}$ is 
a projective family; once we prove this property, the general procedure 
outlined in the previous section gives us the Hilbert space of the 
PL category. 

The only non-trivial property to prove is that the family 
of graphs $\C_{\rm PL}$ is directed. For instance, according to the 
definition of a graph given in last section, the family of all the graphs 
with piecewise smooth edges is not directed. In this case, 
two edges of different graphs can intersect an infinite number of 
times; such two graphs would only accept a common refinement with an 
infinite number of edges, that according to our definition is not a graph. 

We will construct a graph $\c_3$ that refines two given 
graphs $\c_1$ and $\c_2$. 

A trivial property of PL edges lies in the heart of our construction; 
due to its importance, it is stated as a lemma. 
\begin{lemma} \label{finiteness} 
Given two edges of different graphs 
$e_1 \in \c_1$ and $e_2 \in \c_2$, we know that $e_1 \cap e_2$ has 
{\em finitely many} connected components. 
These connected components are either isolated points or piecewise linear 
segments. 
\end{lemma} 

Now we start our construction. 
First we note that every graph $\c$ is refined by a graph 
$\c^\prime$ constructed from $\c$ simply by adding a finite number 
of vertices $v\in V^\prime$ 
in the interior of its edges (and by splitting the edges in the points where 
a new vertex sits). 

Because of lemma~\ref{finiteness}, we know that 
given two graphs $\c_1,\c_2\in C_{\rm PL}$ we can refine each of them 
trivially by adding {\em finitely many} new vertices to form the graphs 
$\c_1^\prime \geq \c_1, \c_2^\prime \geq \c_2$ that satisfy the following 
property. Every edge $e_1\in E_{\c_1^\prime}$ falls into one of the three 
categories given bellow:  
\begin{itemize} 
\item $e_1$ does not intersect any edge of $\c_2^\prime$. \label{1}
\item $e_1$ is also an edge of $\c_2^\prime$; 
$e_1\in E_{\c_2^\prime}$. \label{2}
\item $e_1$ intersects an edge $e_2$ of $\c_2^\prime$ at vertices (one or two) 
of both graphs $e_1 \cap e_2 \subset V_{\c_1^\prime}$, 
$e_1 \cap e_2 \subset V_{\c_2^\prime}$. \label{3}
\end{itemize} 

A direct consequence of these properties is the following: 
\begin{lemma} \label{pl-projective} 
The graph $\c_3$ defined by 
$E_{\c_3}=E_{\c_1^\prime}\cup E_{\c_2^\prime}$ and 
$V_{\c_3}=V_{\c_1^\prime}\cup V_{\c_2^\prime}$ is a refinement of 
$\c_1^\prime$ and $\c_2^\prime$. By the properties of the 
partial ordering relation it follows that $\c_3$ is also a refinement of the 
original graphs $\c_3 \geq \c_1 , \c_3 \geq \c_2$; thus the family of 
piecewise linear graphs $\C_{\rm PL}$ is a projective family. 
\end{lemma} 

In the light of lemma~\ref{pl-projective}, 
the rest of the construction is a simple application 
of the general framework described in the previous section. 
There is a canonical projection $p_\c$ from the space of generalized 
connections $\overline{\cal A}_{\rm PL}$ to the spaces of connections 
${\cal A}_\c$ on graphs $\c \in \C_{\rm PL}$ given by,
\be
p_\c\;\;:\;\overline{\cal A}_{\rm PL} 
\rightarrow 
{\cal A}_\c,\;\;\;p_\c((A_{\c^\prime}) _{\c^\prime\in \C_{\rm PL}})
:=A_\c. 
\ee 
This projective structure is the main ingredient that yields the Hilbert 
space of the connection representation in the PL category. 
Below we state our result concisely. 
\begin{theorem} \label{pl-h} 
The completion (in the sup norm) of the family of 
functions $p_\c^* f_\c(\bar{A})$, defined by graphs $\c \in \C_{\rm PL}$, 
is an Abelian $C^*$ algebra $Cyl_{\rm PL}$. 
A cyclic representation of $Cyl_{\rm PL}$ is provided by the Hilbert space 
\be
{\cal H}_{\rm kin_{PL}}:=L^2(\overline{\cal A}_{\rm PL},d\m_0).
\ee 
that results after completing $Cyl_{\rm PL}$ in the norm provided by the 
Ashtekar-Lewandowski measure $\m_0$. 
\end{theorem}

In the manner described in the previous section we can also consider the 
space of gauge invariant states and obtain 
${\cal H}^{\prime}_{\rm kin_{PL}}$ that is is spanned by spin network states 
labeled by piecewise linear graphs. 

\subsection{The combinatoric category}

In this subsection we introduce the family of combinatoric graphs that 
leads to a manifestly combinatoric approach to quantum gauge theory. 
The construction of combinatoric graphs uses as a corner stone 
the same stone that serves as the combinatoric foundation of topology. 
Thus, our construction provides a quantum/combinatoric 
model for physical space, the space where physical processes take place. 

Simplicial complexes appear first as the combinatoric means of 
capturing the topological information of a topological space $X$. 
By definition, a {\em simplicial complex} $K$ is a set of finite 
sets closed under formation of subsets, formally: 
\be 
x\in K \, \hbox{and} \, y \subset x 
\; \Rightarrow \; y\in K \quad . 
\ee 
A member of a simplicial complex $x\in K$ is called an $n$-simplex if 
it has $n+1$ elements; n is the dimension of $x$. Generically, the set 
of which all simplices are subsets is called the vertex set and denoted 
by $\L$. Some examples of simplicial complexes are given in figure 1 

\hskip 1.7in\epsfbox{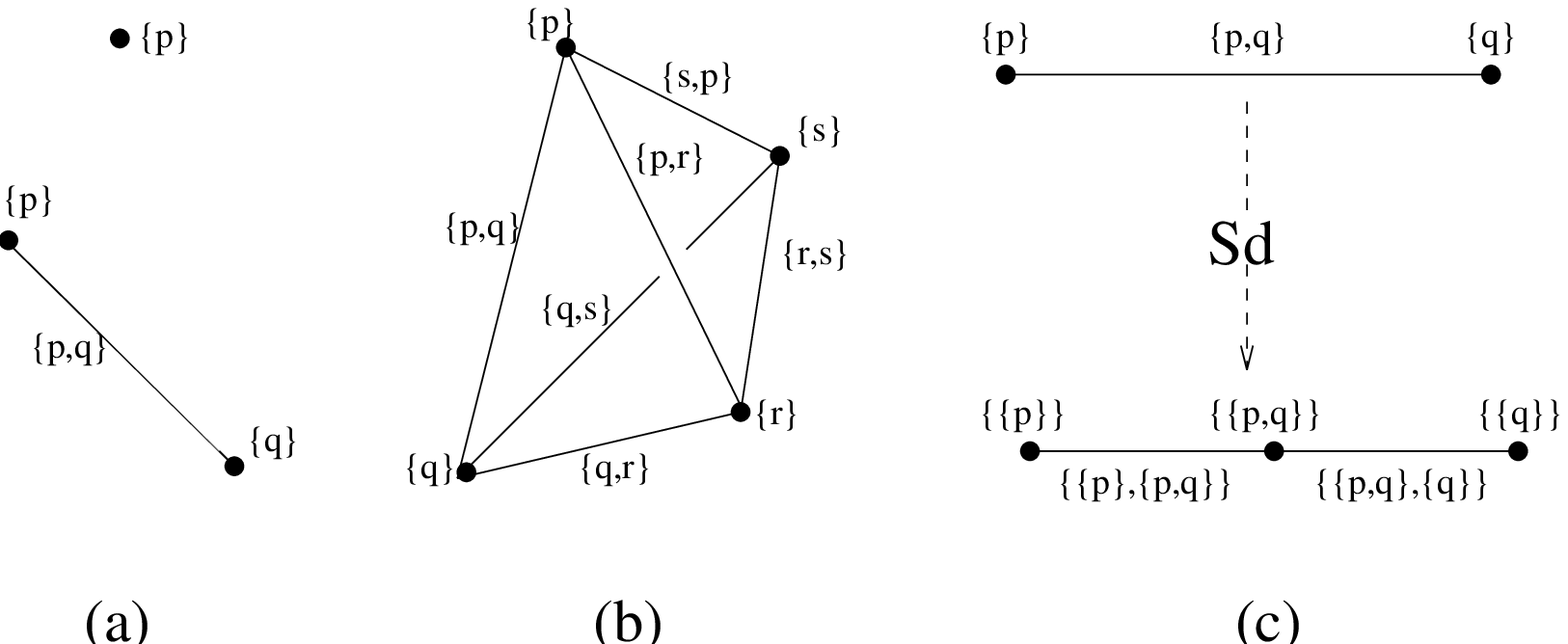} 

\bigskip 

\noindent 
{\small {\bf Fig. 1}
a) Geometrical representations of a zero dimensional simplex $x=\{ p\}$ 
and a one dimensional simplex $x=\{ p,q\}$. 
{\em The simplices are the sets}; in the figures, what we draw 
are the geometric realizations $\D_x$ of the abstract simplices $x$. 
b) A two dimensional complex is a set of simplices of dimension 
smaller or equal to two. In this case the complex 
$K=\{ \{ p\},\{ q\},\{ r\},\{ s\}, 
\{ p, q\},\{ q, r\},$ \\ 
$\{ r, p\},
\{ s, p\},\{ s, q\},\{ s, r\}, 
\{ p, q, r\},\{ p, q, s\},\{ q, r, s\},
\{ r, p, s\},\{ p, q, r, s\} \}$ 
represents a sphere $S^2$. 
Figure (1b) is the geometric realization 
$\| K^1\|$ of the one dimensional subcomplex of $K$ given by 
$K^1=\{ \{ p\},\{ q\},\{ r\},\{ s\}, \{ p, q\},\{ q, r\},\{ r, p\},
\{ s, p\},\{ s, q\},\{ s, r\}$. 
c) The vertices of the baricentric subdivision $Sd(K)$ are the simplices of 
$K$. For example if $K=\{ \{ p\},\{ q\},\{ p, q\} \}$ then 
$Sd(K)=\{ \{ \{ p\} \}, \{ \{ p, q\} \}, \{ \{ q\} \}, 
\{ \{ p\} , \{ p, q\} \}, \{ \{ p, q\}, \{ q\} \}\}$. 
}

Given an open cover ${\cal U}(\L)= \{ U_\l : \l \in \L \}$ 
of a topological space 
$X$ the information about the relative position of the open sets 
$U_1, U_2 , ... \in {\cal U}(\L)$ is the combinatoric information 
that the {\em nerve} $K(\L)$ of ${\cal U}(\L)$ casts. 
The simplicial complex $K(\L)$ 
is the set of all finite subsets of $\L$ such that 
\be 
\bigcap_{\l \in \L} U_\l \neq \emptyset 
\ee 
Using the information encoded in the $K(\L)$ one can often 
recover the topological space $X$. More precisely, every open cover 
${\cal U}(\L)$ of $X$ admits a refinement ${\cal U}^\prime (\L^\prime )$ 
such that the geometric realization (to be defined bellow) 
of its nerve is homeomorphic to $X$, $|K(\L^\prime )| \approx X$. 
This is the sense in which simplicial complexes constitute a combinatoric 
foundation of topology. 

A simplicial complex stores 
topological information combinatorically, but the same 
information can be encoded in a geometric fashion (see 
\cite{fritsch-piccinini}). 
The {\em geometric realization} $|K|$ of a simplicial complex 
$K= K(\L)$, is the subset of $R^{\L}$ given by 
$|K|:= \bigcup_{x \in K} \D_x$ where $\D_x$ is a geometrical simplex 
represented as a segment of a plane of codimension one, embedded in 
$R^{x}$; more precisely, 
\be 
\D_x:= \left\{ \mathbf{s}:=(s_\l : \l \in \x)\in I^x : 
\sum_{\l \in x} s_\l = 1 \right\}
\ee 
where $I=[0,1]$ is the unit interval. The topology of $|K|$ is determined 
by declaring all its geometrical simplices $\D_x$ to be closed sets. 

Our main purpose is to find a combinatoric analog of a generalized 
connection. We need to find the appropriate concept of the space of all 
combinatoric graphs; then a generalized connection will be an assignment of 
group elements to the edges of the graphs. We could fix a simplicial complex 
$K$ to represent the base space and consider that a combinatoric graph is a 
one-dimensional subcomplex $\c \subset K$. The resulting model would 
properly describe topological field theories, but we want to generate 
a model for gauge theories with local degrees of freedom. To achieve 
our goal, we replace physical space (the base space) with a sequence of 
simplicial coplexes $K_0, K_1, \ldots $ that are finer and finer. 

The concept of baricentric subdivision give us the option of generating 
finer and finer simplicial complexes. Given a simplicial complex $K$ its 
{\em baricentric subdivision} $Sd(K)$ is defined as the simplicial complex 
constructed by assigning a vertex to every simplex of $K$, $\L= K$. 
Then, the simplices of $Sd(K)$ are the finite subsets $X\subset \L$ 
that satisfy 
\be 
x,y \in X \Rightarrow x\subset y \; \hbox{or} \; y\subset x 
\ee 
A geometric representation of the operation baricentric subdivision $Sd$ is 
given in figure 1. 

Our approach to quantum gauge theory replaces the base space $\S$ 
with a sequence of simplicial complexes 
$\{K, Sd(K), \ldots , Sd^n(K), \ldots \}$ such that $|K| \approx \S_0$, 
where $\S_0$ is a compact Hausdorff three dimensional manifold. This concept 
of space leads to the definition of combinatoric graphs. 

A {\em combinatoric graph} $\c \in \C_C$ is simply a graph, according to 
the definition given in the previous section, where the set of vertices 
$V_\c$ and the set of edges $E_\c$ are restricted to be subsets of the set 
of points $V(K)$ and the set of oriented paths $E(K)$. 

In the combinatoric representation of space, a point $p\in V(K)$ is 
represented by an equivalence class of sequences of the kind 
$\{ p_n, p_{n+1}=Sd(p_n), p_{n+2}=Sd^2(p_n), \ldots \}$ of 
zero-dimensional simplices $p_n \in Sd^n(K)$, $p_{n+1} \in Sd^{n+1}(K)$, etc. 
Noteably one single element of the sequence determines the 
whole sequence. 
Two sequences $\{ p_n, p_{n+1}= Sd(p_n), \ldots \}$ 
$\{ q_m, q_{m+1}= Sd(q_m), \ldots \}$ 
are equivalent if all their elements coincide, $p_s=q_s \in Sd^s(K)$ 
for all $s\geq \max(n,m)$. 

The definition of oriented paths follows the same idea, but is a little more 
involved. First we will define paths, then oriented paths, and composition 
of oriented paths. 
A path $e\in P(K)$ is an equivalence class of sequences 
$\{ e_n, e_{n+1}=Sd(e_n), \ldots \}$ of one dimensional subcomplexes 
$e_n \subset Sd^n(K)$ such that the geometric realizations of its elements are 
homeomorphic to the unit interval $|e_n| \approx I$. Again, two sequences 
$\{ e_n, e_{n+1}= Sd(e_n), \ldots \}, \{ f_m, \ldots \}$ are equivalent if all 
their elements coincide $e_s=f_s \in Sd^s(K)$ for all $s\geq \max(n,m)$. 

An oriented path 
$e \in E(K)$ is a path $e^\prime \in P(K)$ and a sequence of relations that 
order the vertices%
\footnote{
Here the term vertex refers to a zero-dimensional simplex in the one 
of the one-dimensional subcomplexes $e_n$ in the path $e$. It should not be 
confused with a vertex $v\in V_\c$ of a combinatoric graph. 
} 
of each of the one-dimensional subcomplexes $e_n^\prime$ in the path. 
We denote the initial point of a path by $e(0)\in V(K)$ and 
it is defined by the class of the sequence of initial vertices 
$e(0)=[ \{ e_n(0), e_{n+1}(0)=Sd(e_n(0)), \ldots \} ]\in V(K)$; 
the final point of a combinatoric path is denoted by $e(1)\in V(K)$. 
Composition of two oriented paths 
$e, f \in E(K)$ is possible if they intersect only at the final point of 
the initial path and the initial point of the final path 
$[ \{ e_n\cap f_n, e_{n+1}\cap f_{n+1}, \ldots \} ] = e(1) = f(0)$; 
it is denoted by $f\circ e \in E(K)$ and is defined by 
\be 
f \circ e=[ \{ (f\circ e)_n =f_n \cup e_n, 
(f\circ e)_{n+1}=Sd((f\circ e)_n), \ldots \} ] 
\ee 
and the obvious sequence of ordering relations. 

Given an oriented path $e\in E(K)$ its inverse $e^{-1} \in E(K)$ is defined by 
the same path $e^\prime \in P(K)$ and the opposite orientation. Notice that 
the composition relation is not defined for $e$ and $e^{-1}$; it is possible to 
define combinatoric curves that behave like usual curves, but it is not necessary 
for the purpose of this article. 

Once the set of edges $E$ is endowed with the composition operation, 
the rest of our construction is almost a simple application of 
the general framework reviewed in the previous section. 
The only gap to be filled is proving that the family of combinatoric 
graphs $\C_C$ is directed. 

To prove the directedness in the PL case we used the finiteness property 
stated in lemma~\ref{finiteness}; an adapted statement of 
this same property holds trivially in the combinatoric case. 
\begin{lemma} \label{C-finiteness} 
The intersection of 
two one dimensional subcomplexes $e_n, f_n \subset Sd^n(K)$, defining 
the paths $e, f \in P(K)$ respectively, has 
{\em finitely many} connected components. 
These connected components are either isolated zero-dimensional simplices 
or one-dimensional subcomplexes homeomorphic to the unit interval. 
That is, 
\be 
e_n\cap f_n = (\bigcup_{i=1}^N p(i)_n ) \bigcup (\bigcup_{j=1}^M g(j)_n )
\ee 
where $p(i)_n\subset Sd^n(K)$ is a zero-dimensional simplex and 
$I \approx g(i)_n\subset Sd^n(K)$. In addition, 
$p(i)_n \cap p(j)_n = p(i)_n \cap g(j)_n = g(i)_n \cap g(j)_n = 
\emptyset$ for all $i\neq j$. 

By defining the appropriate notion of union and intersection of classes of 
sequences we can state the result as 
\be 
e\cap f = (\bigcup_{i=1}^N p(i) ) \bigcup (\bigcup_{j=1}^M g(j) )
\ee 
where $p(i)\in V(K)$, $g(j)\in P(K)$, and 
$p(i) \cap p(j) = p(i) \cap g(j) = g(i) \cap g(j) = \emptyset$ for all 
$i\neq j$. 
\end{lemma} 

Therefore, the 
construction of a graph $\c_3\in \C_C$ that refines two given 
graphs $\c_1, \c_2 \in \C_C$ is just an adaptation of the construction 
given for the piecewise linear case. 

Using lemma~\ref{C-finiteness} it is easy to prove that 
given two graphs $\c_1,\c_2\in C_C$ we can refine each of them 
trivially by adding {\em finitely many} new vertices; forming graphs 
$\c_1^\prime \geq \c_1, \c_2^\prime \geq \c_2$ such that every 
edge $e_1\in E_{\c_1^\prime}$ falls in one of the three categories 
(\ref{1}), (\ref{2}), (\ref{3}) itemized in the previous subsection. 

From the previous construction the following lemma is evident. 
\begin{lemma} \label{C-projective} 
Let $\c_3$ be the graph defined by \\
$V_{\c_3}:=V_{\c_1^\prime}\cup V_{\c_2^\prime}\subset V(K)$ and 
$E_{\c_3}:=E_{\c_1^\prime}\cup E_{\c_2^\prime}\subset E(K)$. 

$\c_3$ is a refinement of 
$\c_1^\prime$ and $\c_2^\prime$. By the properties of the 
partial ordering relation it follows that $\c_3$ is also a refinement of the 
original graphs $\c_3 \geq \c_1 , \c_3 \geq \c_2$; thus the family of 
combinatoric graphs $\C_C$ is a projective family. 
\end{lemma} 

Following the general framework described in the previous section 
we will complete the construction of our combinatoric/quantum model 
for gauge theory. 
There is a canonical projection $p_\c$ from the space of generalized 
connections $\overline{\cal A}_C$ to the spaces of connections 
${\cal A}_\c$ on graphs $\c \in \C_C$ given by,
\be
p_\c\;\;:\;\overline{\cal A}_C 
\rightarrow 
{\cal A}_\c,\;\;\;p_\c((A_{\c^\prime}) _{\c^\prime\in \C_C})
:=A_\c. 
\ee 
These projections are the key ingredient that yields the Hilbert 
space of the connection representation in the combinatoric category. 
Below we state our result concisely. 
\begin{theorem} \label{C-h} 
The completion (in the sup norm) of the family of 
functions $p_\c^* f_\c(\bar{A})$, defined by graphs $\c \in \C_C$, 
is an Abelian $C^*$ algebra $Cyl_C$. 
A cyclic representation of $Cyl_C$ is provided by the Hilbert space 
\be
{\cal H}_{\rm kin_C}:=L^2(\overline{\cal A}_C,d\m_0).
\ee 
that results after completing $Cyl_C$ in the norm provided by the 
Ashtekar-Lewandowski measure $\m_0$. 
\end{theorem}

As described in the previous section we can consider the 
space of gauge invariant states and get ${\cal H}^{\prime}_{\rm kin_C}$ 
that is is spanned by spin network states labeled by combinatoric graphs. 

The constructions, given in this and the previous subsection, of the Hilbert 
spaces for the piecewise linear and the cobinatoric categories were 
similar. Despite the parallelism, the resulting Hilbert spaces 
are completely different. A property that marks the difference is the size 
of these Hilbert spaces. 
\begin{theorem} \label{C-separable} 
The Hilbert space 
${\cal H}^{\prime}_{\rm kin_C}$ is separable. 
\end{theorem} 
Proof -- We will prove that the spin network basis 
is countable in the combinatoric case. 

We did not describe precisely the spin network basis, but we stated that 
two spin network states $S^1_{\vec{\c} , j(e), c(v)}(A)$, 
$S^2_{\vec{\d} , j(e), c(v)}(A)$ are
orthogonal if $\c \neq \d$ or if their edge's colors are different. 

Let $L_{\c ,j(e)}$ be the space spanned by all the spin network states 
with labels $\vec{\c}, j(e)$. Our task is bound $n=\dim(L_{\c ,j(e)})$. 
We know that $n$ is less than the number of labels that we would get 
by assigning not one integer but three integers to the graphs edges.  
The first integer $j(e)$ labels the irreducible representation 
assigned to $e$, and the other two $m_L(e), m_R(e)$ determine basis 
vectors in the vector space selected by $j(e)$. With these basis vectors 
sitting at both ends of every edge we can label any set of 
(generally non gauge invariant) contractors for the vertices. 

Thus, the spin network basis is countable if the set of 
finite subsets of 
\be 
E(K) \times \mathbf{N}  
\ee 
is countable. 
Then to prove the theorem we just have to show that the set $E(K)$ 
is countable, which in turn reduces to prove that the set of paths $P(K)$ 
is countable. 

A path $e\in P(K)$ is determined by a sequence of one-dimensional subcomplexes 
that are all related by baricentric subdivision. Therefore, a path $e\in P(K)$ 
can be specified by just one one-dimensional subcomplex of an appropriate 
$Sd^n(K)$. A particular one-dimensional subcomplex can be described by 
specifying which of the one-dimensional and zero-dimensional simplices 
belong to it. We can use the set $\{ 0, 1 \}$ to specify which simplex belong 
or does not belong to a particular subcomlpex. 

Therefore, there is an onto map 
\be 
M: \bigcup_{n=1}^\infty Sd^n(K) \times \{ 0, 1 \} \to P(K) 
\ee 
since a countable union of finite sets is countable and each $Sd^n(K)$ is 
finite, we have proved that $P(K)$ is countable. $\Box$ 

\section{Physical observables and physical states}\label{homeos}

In this section we construct the Hilbert 
space of physical states of our model for quantum gauge theory; 
where we can represent the algebra of physical (gauge and 
``diffeomorphism'' invariant) observables. Our 
quantization procedure follows the same steps as in the analytic 
category; that is, it follows 
(a refined version of) the algebraic quantization 
program \cite{l,alg-quant}. When we deal 
with theories with extra constraints, like gravity, we need to solve these 
extra constraints to find the space of physical states. 

Since the issue of ``diffeomorphism'' invariance acquires quite 
different faces 
in the PL and combinatoric categories, we tackle it first for the PL category. 
Then we find the space of physical states of the combinatoric category and 
prove that it is separable and isomorphic to the space of physical states 
of the PL category. 

\subsection{``diffeomorphism'' invariance in the PL category}

Any operator can be defined by specifying its action on the space of cylindrical 
functions $Cyl$ and then using continuity to extend it to the whole Hilbert 
space ${\cal H}_{\rm kin}$. This is what we did to define the unitary 
operators induced by the gauge symmetry and it is what we will do in this 
section to define quantum ``diffeomorphisms.'' 

Our piecewise linear framework is based on the family of graphs 
$\C_{\rm PL}$ selected by a fixed piecewise linear structure in $\S$. 
Therefore, the role of ``diffeomorphisms'' is played by {\em piecewise linear 
homeomorphisms}. It is important to note that the space of such maps 
can be defined as 
\be 
{\rm Hom}_{\rm PL}(\S):= \left\{ h\in {\rm Hom}(\S) : 
h(\C_{\rm PL})= \C_{\rm PL} \right\} \quad . \label{plh}
\ee 

The unitary operator 
$\hat{U}_h :{\cal H}_{\rm kin_{PL}} \to {\cal H}_{\rm kin_{PL}}$ 
induced by a piecewise linear homeomorphism $h$ is determined by its 
action on cylindrical functions 
\be \label{u} 
\hat{U}_h\cdot\Psi_\c(\bar{A}):=\Psi_{h^{-1}(\c)}(\bar{A}).
\ee 

In contrast with our treatment of gauge invariance, the space of 
diffeomorphism invariant states is not the kernel of any Hermitian operator; 
the reason is that the one-dimensional subgroups of 
the diffeomorphism group induce one-parameter families of unitary 
transformations that are not strongly continuous in our Hilbert space 
\cite{analytic}. Another important difference is that the space of 
``diffeomorphism'' invariant states cannot be made a subspace of the Hilbert 
space ${\cal H}_{\rm kin_{PL}}$, 
the solutions are true distributions, i.e., they lie in 
a subspace of the topological dual of $Cyl_{\rm PL}$. 

A distribution  $\bar\phi \in Cyl_{\rm PL}^*$ is ``diffeomorphism'' 
invariant if 
\be
\bar\phi[\hat{U}_h\circ\psi]=\bar\phi[\psi]\;\;\;\forall\;\; h\in
{\rm Hom}_{\rm PL}(\Sigma)
\;\;\;\;{\rm and}\;\;\;\;\psi\in {\rm Cyl}_{\rm PL} .
\ee 

We can construct such distributions by ``averaging'' over the 
group ${\rm Hom}_{\rm PL}(\S)$. The infinite 
size of ${\rm Hom}_{\rm PL}(\S)$ makes a precise definition of the 
group average procedure very subtle. Here we follow 
the procedure used for the analytic category \cite{analytic}. 

An inner product for the space of solutions is given by the 
same formula that defines the group averaging; therefore, 
a summation over all the elements of ${\rm Hom}_{\rm PL}(\S)$ 
would yield states with infinite 
norm. In this sense, prescribing an adequate definition for the 
averaging over the group ${\rm Hom}_{\rm PL}(\S)$ involves 
``renormalization.'' The issue is resolved once the following two 
observations have been made: 
First, the inner product between two states based on graphs 
$\c , \d\in \C_C$ must be zero unless there is a homeomorphism 
$h_0 \in{\rm Hom}_{\rm PL}(\S)$ such that $\c =h_0 \d$. 
Second, our construction of generalized connections assigns group elements 
to unparametrized edges. Therefore, two homeomorphisms that restricted to 
a graph $\c$ are equal except for a reparametrization of the edges of 
$\c$ should be counted only once in our construction of group averaging 
of states based on graph $\c$. 
Thus, we define a map 
$\e:Cyl_{\rm PL} \to Cyl_{\rm PL}^*$ 
that transforms any given gauge invariant 
cylindrical function into a ``diffeomorphism'' invariant 
distribution. We define 
$\e$ acting on spin network states, then by {\em anti}linearity 
we can extend its action to any gauge invariant cylindrical function. 
Averaging a spin network state 
$S_{\vec{\c} , j(e), c(v)}$ produces a s-knot state 
$s_{[\vec{\c}],j(e),c(v)}=\e(S_{\vec{\c},j(e),c(v)})\in Cyl_{\rm PL}^*$ 
defined by 
\be 
s_{[\vec{\c}] , j(e), c(v)} [S_{\vec{\d}^\prime , j(e), c(v)}] 
:= \d_{[\c] [\d]} a ([\c]) \sum_{[h]\in {\rm GS}(\c)}\langle 
S_{\vec{U_{h\cdot h_0}\c} , j(e), c(v)} |
S_{\vec{\d}^\prime , j(e), c(v)} \rangle \label{d}
\ee 
where $\d_{[\c] [\d]}$ is non vanishing only if there is 
a homeomorphism $h_0 \in {\rm Hom}_{\rm PL}(\S)$ 
that maps $\c$ to $\d$, $a([\c])$ is a normalization parameter, 
and $h\in {\rm Hom}_{\rm PL}(\S)$ is any element in the class of 
$[h]\in {\rm GS}(\c)$. The discrete group ${\rm GS}(\c)$ is the group of 
symmetries of $\c$; i.e. elements of ${\rm GS}(\c)$ are maps between the 
edges of $\c$. The group can be constructed from subgroups of 
${\rm Hom}_{\rm PL}(\S)$ as follows: 
${\rm GS}(\c)= {\rm Iso}(\c)/{\rm TA}(\c)$ where 
${\rm Iso}(\c)$ is the subgroup of 
${\rm Hom}_{\rm PL}(\S)$ 
that maps $\c$ to itself, and the elements of 
${\rm TA}(\c)$ are the ones that preserve all the edges of $\c$ separately. 

{\em The Hilbert space of physical states} ${\cal H}_{\rm diff_{PL}}$ 
is obtained after completing the space spanned by the s-knot states 
$\e(Cyl_{\rm PL})$ 
in the norm provided by the inner product defined by 
\be 
(F,G)=(\e(f),\e(g)):=G[f] \quad .  
\ee 

Define the algebra ${\cal A}_{\rm diff_{PL}}^\prime$ to be the algebra 
of operators on ${\cal H}_{\rm kin_{PL}}$ satisfying the following 
two properties: First, for 
$O \in {\cal A}_{\rm diff_{PL}}^\prime$, both $O$ and $O^\dagger$ 
are defined on $Cyl_{\rm PL}$ and map $Cyl_{\rm PL}$ to itself. 
Second, both $O$ and $O^\dagger$ 
are representable in ${\cal H}_{\rm diff_{PL}}$ by means of 
\be \label{obs} 
r_{PL}(\hat{O}) F = r_{PL}(\hat{O}) \e(f) := 
\e( \hat{O} f)  \quad .
\ee 
${\cal A}_{\rm diff_{PL}}$ 
is the analog of the algebra of weak ``observables.'' 
Different weak observables can be weakly equivalent; in the same way, 
many operators 
of ${\cal A}_{\rm diff_{PL}}^\prime$ are represented by the same 
operator in ${\cal H}_{\rm diff_{PL}}$. For example, 
$r_{PL}(\hat{U}_h)=r_{PL}(1) =1$. We can define the algebra of 
classes of operators of ${\cal A}_{\rm diff_{PL}}^\prime$ 
that are represented 
by the same operator in ${\cal H}_{\rm diff_{PL}}$; this algebra is 
faithfully represented in ${\cal H}_{\rm diff_{PL}}$ and is called 
the algebra of physical operators ${\cal A}_{\rm diff_{PL}}$ 
\cite{alg-quant}. Even more, it is easy to prove that every operator 
on ${\cal H}_{\rm diff_{PL}}$ is in the image of 
$r_{PL}({\cal A}_{\rm diff_{PL}})$. 
The algebra of strong observables (Hermitian operators 
invariant under 
gauge transformations and ``diffeomorphisms'') sits inside of 
${\cal A}_{\rm diff_{PL}}$ (with the commutator as product); 
then it is representable in 
${\cal H}_{\rm diff_{PL}}$ faithfully. 

Since (\ref{obs}) maps any observable to a Hermitian 
operator in ${\cal H}_{\rm diff_{PL}}$, 
this representation implements the reality conditions. In 
particular (when the space manifold is three dimensional and the 
gauge group is $SU(2)$), the construction 
provides a ``quantum Husain-Kucha\v{r} model'' \cite{husain-kuchar}, 
that has local degrees of freedom \cite{analytic}. 

An interesting feature of the quantum 
Husain-Kucha\v{r} model (and of any other diffeomorphism invariant 
quantum gauge theory defined over a compact manifold 
$\S$ with $\dim(\S)= 1,2,3$ following our general framework) 
is that the choice of background structure is 
not reflected in the resulting quantum theory. To be precise, fix a 
piecewise linear structure $PL_0$ on $\S$ and construct the algebra of 
physical operators ${\cal A}_{\rm diff_{PL_0}}$ 
(acting on ${\cal H}_{\rm diff_{PL_0}}$) that it induces. 
Given another piecewise linear structure $PL_1$ on $\S$ and a piecewise 
linear homeomorphism connecting both PL structures 
$h_1: \S_{PL_0}\to \S_{PL_1}$, we get a representation of 
${\cal A}_{\rm diff_{PL_0}}$ in ${\cal H}_{\rm diff_{PL_1}}$ by 
$r_{PL_1} (O)= \hat{U}_{h_1}^{-1} O \hat{U}_{h_1}$. In fact, 
$r_{PL_1}: {\cal A}_{\rm diff_{PL_0}} \to {\cal A}_{\rm diff_{PL_1}}$ is 
onto and it is independent of $h$. Thus we can label the operators of 
${\cal A}_{\rm diff_{PL_1}}$ by the elements of 
${\cal A}_{\rm diff_{PL_0}}$. Using ${\cal A}_{\rm diff_{PL_0}}$ as a 
fiducial abstract algebra, the independence of the background PL structure 
on $\S$ may be stated as follows. 

\begin{theorem} \label{pl-equiv} 
Any piecewise linear structure $PL_1$ on a fixed manifold $\S$ 
of dimension $\dim(\S)= 1,2,3$ defines a representation 
$r_{PL_1}({\cal A}_{\rm diff_{PL_0}})$ of 
${\cal A}_{\rm diff_{PL_0}}$. This representation is independent 
of the piecewise linear structure, in the sense that, given any two 
piecewise linear structures $PL_1$ and $PL_2$ on $\S$, the 
representations $r_{PL_1}({\cal A}_{\rm diff_{PL_0}})$ and 
$r_{PL_2}({\cal A}_{\rm diff_{PL_0}})$ are unitarily equivalent. 
\end{theorem} 
Proof --  In dimensions $\dim(\S)= 1,2,3$
it is known \cite{PLstructures} that 
any two PL structures $PL_i$ and $PL_0$ 
are related by a piecewise linear homeomorphism 
$h_i:\S_{PL_0} \to \S_{PL_i}$. This implies that 
$r_{PL_i}({\cal A}_{\rm diff_{PL_0}})$ defined above is a representation 
of ${\cal A}_{\rm diff_{PL_0}}$. That the representations induced by $PL_1$ 
and $PL_2$ are equivalent is trivial. 
$U_{h_2^{-1}\circ h_1}: {\cal H}_{\rm diff_{PL_1}} \to 
{\cal H}_{\rm diff_{PL_2}}$; that $U_{h_2^{-1}\circ h_1}$ is the required 
unitary map and it induces an algebra isomorphism. $\Box$ 

\subsection{Physical observables and physical states 
in the combinatorial category} \label{Cdinv} 

Now our task is to find the analog of knot-classes of combinatoric 
graphs. 
In section~\ref{p-lspaces} we reviewed how is that a simplicial complex 
$K$ encodes combinatorially topological information, and how this 
information can be displayed in its geometric realization $|K|$. 
Then, to decide whether 
or not two combinatoric graphs $\c, \d \in \C_C$ belong to the same 
knot-class we are going to display them in the same space and compare them. 

To this end, we fix the sequence of piecewise linear maps 
\be 
M_n : |Sd^n(K)| \to |K| 
\ee 
defined by successive application of 
the canonical map $M_1: |Sd(K)| \to |K|$ that maps the vertices of 
$|Sd(K)|$ to the baricenter of the corresponding simplex in $|K|$. 
Then, we map every every representative 
$\{ \c_n, c_{n+1}=Sd(\c_n), \ldots \}$ 
of the combinatoric graph $\c$ in to a sequence 
\be 
\{ M_n(|\c_n|), M_{n+1}(|c_{n+1}|)=M_n(|\c_n|), \ldots \}
\ee 
that assigns the same geometric graph $|\c|:= M_n(|\c_n|)$ to every integer. 
Using these maps we are going to define that the combinatoric graphs 
$\c, \d \in \C_C$ are ``diffeomorphic'' if the their corresponding 
geometrical graphs $|\c|, |\d|$ are related by a piecewise linear 
homeomorphism. 

One method in implementing the above idea is to use the sequence of 
maps $M_n$ to induce a map that links the kinematical Hilbert spaces of 
the combinatoric and PL categories. The map 
$M: Cyl_C \to Cyl_{\rm PL}$ is defined by 
\be 
M (f_\c ):= f_{M_n(|\c_n|)}  = f_{|\c|}  \quad . 
\ee 
Now the map $\e:Cyl_{\rm PL}(\amodg) \to Cyl_{\rm PL}^*(\amodg)$ 
induces a new map 
$\e_C:Cyl_C(\amodg) \to Cyl_C^*(\amodg)$
\be 
\e_C := M^* \circ \e \circ M : Cyl_C \to Cyl_C^*
\ee 
that produces ``diffeomorphism'' invariant distributions in the combinatoric 
category. 
Again, we characterize the averaging map by the s-knot states 
$s_{[\vec{\c}]_C , j(e), c(v)} \in Cyl_C^*$ induced by the 
combinatoric spin network states $S_{\vec{\c} , j(e), c(v)}$ 
\be  \label{s_C} 
s_{[\vec{\c}]_C , j(e), c(v)} [S_{\vec{\d}^\prime , j(e), c(v)}] = 
\e_C (S_{\vec{\c} , j(e), c(v)}) [S_{\vec{\d}^\prime , j(e), c(v)}] := 
s_{[\vec{| \c |}] , j(e), c(v)} [S_{\vec{| \d |}^\prime , j(e), c(v)}]
\ee 
As follows from the above formula, the label $[\vec{\c}]_C$ of the s-knot 
states is an equivalence class of oriented combinatoric graphs, where  
$\vec{\c}$ and $\vec{\d}$ are considered equivalent if there is 
$h \in {\rm Hom}_{\rm PL}(|K|)$ such that $h(\vec{| \c |})=\vec{| \d |}$. 

Just as in the PL case, {\em the Hilbert space of physical states} 
${\cal H}_{\rm diff_C}$ 
is obtained after completing the space spanned by the s-knot states 
$\e_C(Cyl_C(\amodg))$ 
in the norm provided by the inner product defined by 
\be 
(F,G)=(\e_C(f),\e_C(g)):=G[f] \quad .  
\ee 

It may seem odd that we are constructing the space of ``diffeomorphism'' 
states without a family of unitary maps called ``diffeomorphisms''. 
The reason for this 
peculiarity is behind the very beginning of our construction. We chose to 
represent space combinatorially with a sequence generated by the simplicial 
complex $K$, and we did not consider the sequence generated by other complex, 
say $L$, even if it had the same topological information $|K| \approx |L|$. 
If we had done that, we would have ended with a kinematical Hilbert space that 
would be made of two copies of the one that we defined here, and these two 
copies would be linked by ``diffeomorphisms''. What we did was 
to construct every thing above the minimal kinematical Hilbert space. A 
relevant question is if by shrinking the kinematical Hilbert space we also 
shrank the space of physical states. Below, we will prove that this is not 
the case. 

Now we state two important characteristics of the spaces of physical states 
of the combinatoric and PL models. 

First, we constructed the space 
${\cal H}_{\rm diff_C}$ using the map $\e_C$; the same 
map can be restricted to give an onto map from the spin network basis of 
${\cal H}^{\prime}_{\rm kin_C}$ 
to the basis of ${\cal H}_{\rm diff_C}$. Since the 
kinematical Hilbert space is separable, we have the following physically 
interesting result. 
\begin{theorem} \label{d-separable}
The Hilbert space ${\cal H}_{\rm diff_C}$ is separable. 
\end{theorem} 

Second, the map $M^*: Cyl_{\rm PL}^* \to Cyl_C^*$ 
can be extended by continuity to link the spaces of 
physical states of the PL and combinatoric categories. 
Using this map we can compare these two spaces. 
\begin{theorem} \label{iso} 
The spaces of physical states in the PL and combinatoric categories 
are naturally isomorphic, 
${\cal H}_{\rm diff_{PL}}\approx {\cal H}_{\rm diff_C}$. 
\end{theorem} 
Proof -- If $\vec{\c_{\rm PL}} = \vec{| \c |}$ 
then $M^*$ identifies the s-knot 
states that they generate by averaging, in other words, 
$ M^* ( s_{[\vec{| \c_{\rm PL} |}] , j(e), c(v)}) = 
s_{[\vec{\c}]_C , j(e), c(v)} $. From the definition of the inner products 
and the definition of the combinatoric s-knot states it follows immediately 
that $M^*$ is an isometry. 

Since the spaces of physical states were 
constructed by completing the vector spaces spanned by the s-knot states, 
the theorem is a consequence of the following lemma, which will be proved in 
the appendix. 
\begin{lemma} \label{Cknots} 
In any knot-class of PL oriented graphs $[\vec{\c_{PL}}]$ there is at 
least one representative that comes from the geometric representation of a 
combinatoric oriented graph $\vec{| \c |} \in [\vec{\c_{PL}}]$. 
\end{lemma} 
$\Box$ 

Now we proceed to construct a representation of the algebra of physical 
operators in the combinatoric category. 
As in the PL category, we 
define the algebra ${\cal A}_{\rm diff_C}^\prime$ to be the algebra 
of operators on ${\cal H}_{\rm kin_C}$ that satisfy the following 
two conditions: First, for 
$O \in {\cal A}_{\rm diff_C}^\prime$, both $O$ and $O^\dagger$ 
are defined on $Cyl_{\rm C}$ and map $Cyl_{\rm C}$ to itself. 
Second, both $O$ and $O^\dagger$ 
are representable in ${\cal H}_{\rm diff_C}$ by means of 
\be \label{obsC} 
r_C(\hat{O}) F = r_C(\hat{O}) \e_C(f) := 
\e_C( \hat{O} f)  \quad .
\ee 
We are interested in the algebra of 
classes of operators of ${\cal A}_{\rm diff_C}^\prime$ 
that are represented 
by the same operator in ${\cal H}_{\rm diff_C}$; this algebra is 
faithfully represented in ${\cal H}_{\rm diff_C}$ and is called 
the algebra of physical operators ${\cal A}_{\rm diff_C}$ 
\cite{alg-quant}. 
In contrast with the PL case, in the combinatoric framework the 
``diffeomorphism group'' does not have a natural action; for this 
reason the notion of strong observables can not be intrinsically 
defined. However, it is easy to prove that in the PL case 
the subset of ${\cal A}_{\rm diff_{PL}}$ consisting of Hermitian 
operators is, in fact, the algebra of strong observables 
(with the commutator as product). Therefore, in the combinatoric 
category we can regard the algebra of Hermitian operators 
in ${\cal A}_{\rm diff_C}$ as the algebra of strong observables; 
this algebra is naturally represented in ${\cal H}_{\rm diff_C}$. 

Since (\ref{obsC}) maps any observable to a Hermitian 
operator in ${\cal H}_{\rm diff_C}$, 
this representation implements the reality conditions. In 
particular (when the space manifold is three dimensional and the 
gauge group is $SU(2)$), 
the construction provides another ``quantum 
Husain-Kucha\v{r} model'' \cite{husain-kuchar}.  A natural 
question is whether the PL and combinatoric models are physically 
equivalent or not. We saw that the 
algebra ${\cal A}_{\rm diff_{C(K)}}$ is represented in 
${\cal H}_{\rm diff_C(K)}$ by $r_{C(K)}$; it is also natural to 
give the representation $d_K({\cal A}_{\rm diff_{C(K)}})$ on 
${\cal H}_{\rm diff_{PL(|K|)}}$ by 
$d_K(\hat{O}) F_{PL} = d_K(\hat{O}) (\e \circ M f_C) := 
\e \circ M (\hat{O} f_C)$. This two representations are identified by 
the isomorphism exhibited in (\ref{iso}), more precisely: 
\begin{theorem} \label{equiv} 
The representations $r_{C(K)}({\cal A}_{\rm diff_{C(K)}})$ 
on ${\cal H}_{\rm diff_C(K)}$ and 
$d_K({\cal A}_{\rm diff_{C(K)}})$ on ${\cal H}_{\rm diff_{PL(|K|)}}$
of the algebra ${\cal A}_{\rm diff_C(K)}$ are unitarily equivalent. 
In addition if $\dim(\S)= 1,2,3$
this algebra does not depend on $K$ but only on the topology 
of $|K|\approx \S$; 
the combinatoric and PL frameworks (based on 
the choice of the Ashtekar-Lewandowski measure 
$\m_0$ on ${\cal H}_{\rm kin}$) 
provide unitarily equivalent 
representations of the abstract algebra ${\cal A}_{\rm diff_{\S}}$. 
\end{theorem} 
Proof -- The unitary equivalence of 
$r_{C(K)}({\cal A}_{\rm diff_{C(K)}})$ and 
$d_K({\cal A}_{\rm diff_{C(K)}})$ is given by the unitary map 
$M^*: {\cal H}_{\rm diff_{PL(|K|)}} \to {\cal H}_{\rm diff_C(K)}$. 

$d_K({\cal A}_{\rm diff_{C(K)}})$ maps ${\cal A}_{\rm diff_C(K)}$ 
onto the algebra of operators on ${\cal H}_{\rm diff_{PL(|K|)}}$ and 
the representation is faithful; the same thing happens for 
the combinatoric model based on a different simplicial complex $L$. 
From theorem (\ref{pl-equiv}) we know that if $\dim(\S)= 1,2,3$ 
for any two simplicial 
complexes $K,L$ such that $|K|\approx |L|\approx \S$ the Hilbert spaces 
${\cal H}_{\rm diff_{PL(|K|)}}$, ${\cal H}_{\rm diff_{PL(|L|)}}$ and 
the algebras of operators on them 
are {\em identified} (unambiguously) by a unitary map. Since 
$d_K({\cal A}_{\rm diff_{C(K)}})$, 
$d_L({\cal A}_{\rm diff_{C(L)}})$,
$d_K({\cal A}_{\rm diff_{PL(|K|)}})$ and 
$d_L({\cal A}_{\rm diff_{PL(|L|)}})$ 
label the operators on 
${\cal H}_{\rm diff_{PL(\S)}}$, there is an unambiguous 
invertible map {\em identifying} these algebras. Thus the family of 
all these equivalent algebras may be regarded as the abstract algebra 
${\cal A}_{\rm diff_{\S}}$ and the combinatoric and PL frameworks are 
procedures that yield unitarily equivalent 
representations of this abstract algebra. $\Box$ 

From the theorems it follows that the PL and combinatoric frameworks are 
physically equivalent. They yield representations of the  algebra of 
physical observables in separable Hilbert spaces; hence, maintaining the
usual interpretation of quantum field theory \cite{wightman}. 

\section{Discussion and Comparison} \label{compari} 

In this paper we have presented two models for quantum gauge field theory. 
We proved that the two models represented the algebra of physical observables 
in separable Hilbert spaces ${\cal H}_{\rm diff_{PL}}$ and 
${\cal H}_{\rm diff_C}$; furthermore, we proved that the two models 
where physically equivalent in the sense that they 
gave rise to unitarily equivalent representations of the algebra of 
physical observables. 
The equivalence of the two models is a good feature, but we may still 
ask if by choosing a different background structure (like a different 
PL structure for our base space manifold) we could have arrived at a 
physically different model. 
In contrast to the analytic case, this problem has been thoroughly studied 
(see for example \cite{PLstructures}). 
For example, in dimensions $\dim(\S) =1,2,3$ any two 
PL structures, like any two differential structures, 
of a fixed topological manifold $\S$ are known to be equivalent 
in the sense that they are related by a PL homeomorphism (diffeomorphism). 
Then, if the base space is three dimensional (like in canonical 
quantum gravity) all the different choices of background structure would 
yield unitarily equivalent representations of the algebra of physical 
(gauge and diffeomorphism invariant) 
observables (the unitary map given by a quantum ``diffeomorphism''). 

Our quantum models are not equivalent to the ones created in the analytic 
category \cite{analytic}; for instance, in the analytic category the 
physical Hilbert space is not separable. The reason for this size difference 
is not that the family of piecewise analytic graphs is too big; the kinematic 
Hilbert space of the PL category is also not separable. In a separate 
paper \cite{CSfromLQG} we show that concept of knot-classes that should be 
used in the piecewise 
analytic category is with respect to the group of maps defined 
by 
\be 
{\rm Pdiff}_\o (\S):= \left\{ h\in {\rm Hom}(\S) : 
h(\C_\o)= \C_\o \right\} \quad . \label{pw}
\ee 
In the appendix we show how to adapt the proof of lemma~\ref{Cknots} 
to show that every (modified) knot-class of piecewise analytic graphs 
has a representative induced by a combinatorial graph. Then, 
theorem~\ref{iso} and theorem~\ref{equiv} 
have analogs proving that the the Hilbert space of physical states of the 
piecewise analytic category is also separable and that 
the representation of the algebra of physical observables 
given by the piecewise analytic 
category is unitarily equivalent to the one provided by the combinatorial 
framework. 

We can expect (the author does) 
that a more satisfactory understanding of field theory 
may arise from this combinatorial picture of quantum geometry. 
The bridge between three-dimensional quantum geometry and a smooth 
macroscopic space-time is the missing ingredient to complete this picture 
of quantum field theory. Three unsolved problems prevent us from building 
this bridge. Dynamics in quantum gravity is only partially understood 
\cite{thomas-lee-rueman}. The emergence of a four-dimensional picture 
from solutions to the constraints has just begun to be explored 
\cite{carlo-mike}. And the statistical mechanics needed to find the 
semi-classical/macroscopic behavior of the theory of 
quantum geometry is also at its developing stage \cite{weeves-kiril}.

\section*{Acknowledgments} 
This paper greatly benefited from the wonderful 
course on QG given by Abhay Ashtekar on the fall of 95, as well as from 
many discussions with Abhay Ashtekar and Alejandro Corichi. 
I also need to acknowledge illuminating conversations, suggestions and 
encouragement from John Backer, John Baez, 
Chris Beetle, Louis Crane, Kiril Krasnov, Seth Major, 
Guillermo Mena Murgain, 
Jorge Pullin, Carlo Rovelli, Lee Smolin, Madhavan Varadarajan, Thomas 
Thiemann, and the editorial help of Mary-Ann Hall. 
Support was provided by Universidad Nacional Aut\'onoma de M\'exico (DGAPA), 
and grants NSF-PHY-9423950, NSF-PHY-9396246, research funds of the Pennsylvania 
State University, the Eberly Family research fund at PSU and the Alfred P. 
Sloan foundation. 

\section*{Appendix} 
First we will prove lemma~\ref{Cknots}, and then, indicate how the proof 
can be extended to link our models and the refined version of the analytic 
category that was mentioned in section~\ref{compari}. 

Given an oriented PL graph $\vec{\c}_{PL} \subset |K|$ we will construct 
an oriented combinatoric graph $\vec{\c}$ and a piecewise 
linear homeomorphism 
(PL map) $h: |K| \to |K|$ such that $h(|\vec{\c}|) = \vec{\c}_{PL}$. 
The construction has four steps. 
\begin{enumerate} 
\item \label{0}
Let $\vec{\c}_{PL}^\prime$ be a refinement of $\vec{\c}_{PL}$ such that 
for every $\D(x_n) \in |K|$ $e \in E_{\c_{PL}^\prime}$ implies that 
$e \cap \D(x_n)$ is empty or 
linear according to the affine coordinates given by $\D(x_n)$. 
\item \label{11}
Find $n$ 
such that $M_n(|Sd^n(K)|)$ separates the vertices of 
$\c_{PL}^\prime$ to 
lie in different geometric simplices $M_n(\D(x_n))$, where 
$\D(x_n) \in |Sd^n(K)|$. Namely, we chose $n$ as big as necessary to 
accomplish a fine enough refinement of $|K|$, where 
$v_1, v_2 \in M_{n}(\D(x_n))$ for 
two different vertices of the PL graph 
$v_1, v_2 \in V_{\c_{PL}^\prime}$ does not 
happen. 

\item \label{12}
Let $h_1:|K| \to |K|$ be the PL map that fixes the vertices of 
$M_n(|Sd^n(K)|)$ and sends the new vertices $M_{n+1}(v(\D(x_n)))$ 
of $M_{n+1}|(|Sd^{n+1}(K)|)$ to 
\begin{enumerate} 
\item $v\in V_{\c_{PL}^\prime}$ if $v$ lies in the 
interior of $M_n(\D(x_n))$; 
symbolically, $v \in (M_n(\D(x_n)))^{\circ}$. 
\item the baricenter of $M_n(\D(x_n))$ if 
there is no $v\in V_{\c_{PL}^\prime}$ such that 
$v \in (M_n(\D(x_n)))^{\circ}$. 
\end{enumerate} 

\item \label{13}
Find $m$ such that 
$h_1(M_{n+m}(|Sd^{n+m}(K)|))$ separates the edges of $\c_{PL}$. 
Stated formally, find $m\geq 1$ such that 
$\c_{PL}\cap h_1(M_{n+m}(\D(x_{n+m})))^{\circ}$ 
has one connected component or it is empty. 

\item \label{14}
Let $h=h_2\circ h_1:|K| \to |K|$, where $h_2$ is 
the PL map that fixes the vertices of 
$h_1(M_{n+m}(|Sd^{n+m}(K)|))$ and sends the new vertices 
$h_1(M_{n+m+1}(v(\D(x_{n+m}))))$ 
of $h_1(M_{n+m+1}|(|Sd^{n+m+1}(K)|))$ to 
\begin{enumerate} 
\item the baricenter of 
$\c_{PL}\cap h_1(M_{n+m}(\D(x_{n+m})))$ if 
$\c_{PL}\cap (h_1(M_{n+m}(\D(x_{n+m}))))^{\circ}\neq \emptyset$. 
\item the baricenter of 
$h_1(M_{n+m}(\D(x_{n+m})))$ if 
$\c_{PL}\cap (h_1(M_{n+m}(\D(x_{n+m}))))^{\circ}= \emptyset$. 
\end{enumerate} 
\end{enumerate} 

From the construction of $h\circ M_{n+m} : |Sd^{n+m}(K)| \to |K|$ 
it is immediate that $(h\circ M_{n+m})^{-1} (\c_{PL}) = |\c_{n+m}|$ if 
$\c_{n+m}\subset Sd^{n+m}(K)$ is defined by 
\begin{itemize} 
\item The zero-dimesional simplex $p \in Sd^{n+m}(K)$ belongs to 
$\c_{n+m}$ if $(h\circ M_{n+m})^{-1} (\c_{PL}) \cap |p| \neq \emptyset$. 
\item The one-dimesional simplex $e \in Sd^{n+m}(K)$ belongs to 
$\c_{n+m}$ if 
$(h\circ M_{n+m})^{-1}(\c_{PL})\cap |e|^{\circ} \neq \emptyset$. 
\end{itemize} 
Then the obvious orientation of $\c_{n+m}$ defines the oriented 
combinatoric graph $\vec{\c}$ and the pair 
$h$, $\vec{\c}$ satisfies 
\be h(|\vec{\c}|) = \vec{\c}_{PL} 
\ee 
$\Box$ 

To link the combinatoric and the analytic categories we need to fix a 
map $N_0: |K| \to \S_{P\o}$ that assigns a piecewise 
analytic curve in $\S_{P\o}$ to every PL curve of $|K|$. 
Then the map $N: Cyl_C \to Cyl_\o$ defined by 
\be 
N (f_\c ):= f_{N_0 \circ M_n(|\c_n|)}  = f_{N_0(|\c|)} 
\ee 
links the kinematical Hilbert spaces, and 
the map $N^*: Cyl_\o^* \to Cyl_C^*$ links the spaces of 
physical states of the analytic and combinatoric categories. 
As it was argued in section~\ref{homeos} $N^*$ is an isometry between 
${\cal H}_{\rm diff_\o}$ and ${\cal H}_{\rm diff_C}$, which means that 
the two Hilbert spaces are isomorphic if every knot-class of piecewise 
analytic graphs $[\c_\o]$ has at least one representative that 
comes from a combinatoric graph $N_0(| \c |) \in [\c_\o]$. 

An extension of the lemma proved in this appendix solves the issue. 
Given a piecewise analytic 
graph $\c_\o \subset \S_{P\o}$ we can construct 
a combinatoric graph $\c$ and a piecewise analytic map $\f$ such that 
$\f \circ N_0 (|\c|) = \c_\o$. First find a refinement 
$\c_\o^{\prime}$ of $\c_\o$ such that its edges are analytic according to 
the domains of analycity of $\S_{P\o}$. Then, 
define a graph in $|K|$ by 
$\a = N_0^{-1}(\c_\o^{\prime})$ and do steps (\ref{11}), (\ref{12}) 
and (\ref{13}) using $\a$ instead of $\c_{PL}$. 
At this moment $N_0 \circ h_1 \circ M_{n+m}(|Sd^{n+m}(K)|)$ separates 
the edges of $\c_\o^{\prime}$; we only need to find a replacement for 
step (\ref{14}). Our strategy is to find a map of the form 
$\f= \f_2 \circ N_0 \circ h_1$ to solve the problem. 
This would be achieved if the 
piecewise analytic diffeomorphism $\f_2$ fixes the mesh given by 
$N_0 \circ h_1 \circ M_{n+m}(|Sd^{n+m}(K)|)$ and at the same time 
matches the mesh given by 
$N_0 \circ h_1 \circ M_{n+m+1}(|Sd^{n+m+1}(K)|)$ and the graph 
$\c_\o^{\prime}$. The map $\f_2$ needs to send every cell 
$N_0 \circ h_1 \circ M_{n+m}(\D(x_{n+m}))$ to itself and 
matche the graph with analytic edges. An explicit construction 
would be cumbersome, but the existence of such a 
{\em piecewise} analytic map is clear. After this is completed, the 
construction of the combinatoric graph follows the instructions 
given above to link the combinatoric and PL categories.

\end{document}